\documentclass[12pt]{article}
\usepackage[tmargin=1in,bmargin=1in,lmargin=1.25in,rmargin=1.25in]{geometry}

\usepackage{setspace}
\usepackage{csquotes}
\usepackage{kpfonts}
\usepackage{amsmath}
\usepackage{amsfonts}
\usepackage{stmaryrd}
\usepackage{dsfont}
\usepackage{graphicx}
\usepackage[title]{appendix}
\usepackage[font=small,labelfont=bf]{caption}
\usepackage{lipsum}
\usepackage{mwe}
\graphicspath{ {images/} }

\usepackage[]{tocloft}
\addtocontents{toc}{\cftpagenumbersoff{section}}
\addtocontents{toc}{\cftpagenumbersoff{subsection}}

\makeatletter
\renewcommand{\@cftmaketoctitle}{}
\makeatother

\title{Why the Weyl Tile Argument is Wrong}

\date{Lu Chen }
\author{Forthcoming in \textit{British Journal for Philosophy of Science}}
\begin{document}

	\maketitle

	\begin{abstract}
	Weyl famously argued that if space were discrete, then Euclidean geometry could not hold even approximately. Since then, many philosophers have responded to this argument by advancing alternative accounts of discrete geometry that recover approximately Euclidean space. However, they have missed an importantly flawed assumption in Weyl's argument: physical geometry is determined by fundamental spacetime structures independently from dynamical laws. In this paper, I aim to show its falsity through two rigorous examples: random walks in statistical physics and quantum mechanics. 
	
\end{abstract}

%\mbox{}
%\tableofcontents
%\mbox{}
	
\section{Introduction}
		Space (or spacetime) is called `discrete' if it is composed of extended indivisible regions---call them `tiles'. Weyl ([1949]) famously argued that if space were discrete, then Euclidean geometry would not hold even approximately, which would contradict our observations. Therefore, space is not discrete. More specifically, if space were composed of tiles, then the diagonal of a square region would be `equal in length to the side' (Weyl [1949], p.43), which would radically violate the Pythagorean theorem.  It makes no difference how big the region is: even at the macrolevel, the diagonal of a square region would still be equal to its side. Thus space would not be approximately Euclidean at any scale.  As I pointed out in (Chen [2021]), this argument relies on the implicit assumption that the distance between any two tiles is equal to the number of tiles between them, which is a standard assumption about discrete space (for example, see Riemann [1866]).
		
			Since then, many philosophers have proposed solutions to this argument (for example, see van Bendegem [1987], [1995], Forrest [1995], Chen [2021]).\footnote{Note that the Weyl tile argument cannot be solved by simply replacing the square tile arrangement with other simple regular arrangements---Fritz ([2013]) shows that in order to approximate Euclidean geometry at the large scale, the arrangement of the tiles need to be very irregular and complicated.}  While these authors have successfully suggested some alternative accounts of geometry that allow for discrete space, they have missed an importantly flawed assumption in Weyl's argument: physical geometry is determined by fundamental spacetime structures independently from the dynamical laws.\footnote{While this paper is not concerned with evaluating the existent solutions to the argument, I can very briefly talk about some of their weaknesses. The solution I proposed in (Chen [2021]), which I argued to be an improvement over Forrest's ([1995]), relies on a complicated and perhaps contrived account of discrete space (in that account, real-valued primitive distances are held by a vast number of `neighbouring' tiles). Van Bendegem ([1995]) discussed and criticized his earlier ([1987]) proposal, but his new proposal has a complicated ideology: he posited both a set of microscopic geometric entities (`$t$-point', `$t$-line', and so on) and macroscopic ones (`point', `line', and so on), and the latter are governed by principles not reducible to those of the former.
			
	%	In case the transition of the views from my earlier paper to this one is confusing to the readers, let me explain the transition more explicitly. In fact, I was interested in exploring both a dynamical solution and a geometric one to Weyl's argument at the time, but did not have the resources to pursue the direction that this paper takes.
}
	 In this paper, I aim to show its falsity through two rigorous examples: random walks in statistical physics and quantum mechanics.  These examples are intended to be a proof of concept for the claim that physical geometry arises from dynamical laws that do not assume any metric notion (`metric' is a technical notion that generalizes the notion of distances). Even if space is discrete, the right dynamical laws could make space appear approximately Euclidean. This is why Weyl's argument fails.

	In the random walk case, I show that in a two-dimensional discrete space represented by $\mathbb{Z}^2$ (pairs of integers), the probability distribution of a wandering tiny cat starting at any given position showing up at each tile is approximately rotationally invariant (it's called `random walk' because, for any time step, the cat randomly walks from a tile to one of its neighbouring tiles). If such probabilities are the only observable quantities, then we have an embedding map from  the discrete space to Euclidean space that approximately preserves all structures and observations. This means that Euclidean geometry is approximately recovered at the empirical level. 
	
	In the case of quantum mechanics, I show that for a quantum mechanical system starting with a sufficiently spread-out position wavefunction, the amplitude of it reaching each tile at a later time is approximately rotationally invariant. Assuming that the probability of a quantum mechanical system in a region at a time (which is the square of the corresponding amplitude) is the only kind of observational quantities, this again means that Euclidean geometry is approximately recovered at the observational level. The random walk case is chosen for its rigor and conceptual clarity while the case of quantum mechanics adds more physical relevance. 
	
	Based on the two cases, I criticize two `geometricist' assumptions  underlying Weyl's argument (geometricism is roughly the view that geometry is more fundamental than dynamics; the opposite view is called `dynamicism'): (1)  large-scale or observable physical geometry is determined by fundamental spacetime structures independent of dynamical laws; (2) geometric structures including the metric structure are ontologically and explanatorily prior to dynamics and must be presumed by the latter.  Against (1), in the two cases, the dynamical laws play an essential role in determining observational geometry. Against (2), no metric notion is presupposed by the dynamical laws. It is also important to note that, while the two cases are simple toy examples, they are not contrived or \textit{ad hoc}, and therefore suggest a realistic possibility that our actual fundamental laws similarly give rise to Euclidean geometry on the ordinary scales.   Thus, we should reject the geometricist assumptions in our understanding of spacetime. While this paper focuses on criticizing Weyl's argument and thereby defending the possibility of discrete space, it also seeks to draw a general lesson about physical geometry by rejecting the geometricist assumptions.\footnote{In this paper, I will exclusively focus on recovering approximate Euclidean geometry from discrete space in the non-relativistic context. I will not be concerned with the task of recovering approximate Minkowski (chrono)geometry from discrete spacetime, which is vastly more complicated. Nevertheless, the paper can still be suggestive towards discrete spacetime by opening up more possibilities.} (For more discussions on dynamicism, see for example Brown [2005] and Norton [2008].)
	
	The paper (in particular, Section 2 and 3) is technical in nature, but the main text aims to be generally accessible for interested philosophers while the appendices contain all the mathematical details. (The appendices are in fact just as important a part of the paper as the main text.) 
	
	\section{Weyl's Tile Space}
		
		The Weyl tile argument against discrete space implicitly assumes a simple \textit{counting} account of distance in the tile space (Chen [2021]). To explain, we first need to define a topology on the tile space through the primitive notion of `connectedness', which refers to an irreflexive and symmetric binary relation (see Roeper [1995] for a topological framework based on connectedness). For example, for the two-dimensional discrete space, we can postulate the following topology:
		
				 \begin{quote}
			\textsc{Connectedness.} Let the tiles be represented by members of  $\mathbb{Z}^2$. For any tiles $x$ and $y$, they are \textit{connected} iff $|x-y|=1$.\footnote{For $x=(x_1,x_2)$ and $y=(y_1,y_2)$, $|x-y|=\sqrt{(x_1-y_1)^2+(x_2-y_2)^2}$. 
			
		Note that such a topology defines `discrete space' when there is no fundamental metric structure. The notion of `extended indivisible regions' used in defining an intuitive notion of `discrete space' is no longer clear in the lack of a metric. Instead, discrete space can be defined by its topology: for example, every indivisible region is connected to a finite number of indivisible regions. Thanks to a participant at the 2022 APA pacific division meeting for pointing this out.}
			
		\end{quote}   
		 Informally, this says that every tile is connected to its neighbouring four tiles (namely its left, right, up, and down tiles).\footnote{Note that in Weyl's argument, each tile is connected to eight neighbouring tiles rather than four. But the difference does not affect the essence of Weyl's argument.} Then, we can obtain distances between tiles by counting connected tiles between them:
		 
		\begin{quote}
			\textsc{The Counting Account.} For any tiles $A,B$, if $C_1,C_2,...C_n$ are the least number of tiles that are pairwise connected and connected to $A,B$, then the distance between $A,B$ is $n+1$. 
		\end{quote}
	The counting account is the discrete version of the standard path-dependent account of distance, namely that the distance between two points equals the length of a shortest path (or extremal ones in the case of spacetime). In the discrete case, a path is composed of connected tiles and its length equals the number of those tiles. 	The counting account is very intuitive and natural for discrete space and has indeed been endorsed by many, including Riemann in his foundational work ([1866]) for differential geometry. Apriorily, this seems to be the best account due to its simplicity and elegance (Forrest [1995]). 
	
	But the appearance is wrong, because the counting account falsely assumes that physical geometry and the distances we observe  can be determined by fundamental spatial structures independently from dynamical laws that govern how matters behave and interact. As I will argue, this assumption is unwarranted.  (Note that the notion of distance in the counting account and the geometry it determines should be empirical, that is, being observable and measurable by  devices like rigid rods and light rays, since a non-empirical geometry could not contradict our observations as Weyl's argument goes.)
	
	In the upcoming sections, I will offer two rigorous cases in which the underlying spatial structure is as simple as Weyl imagined (minus any metric structure), and yet the apparent  geometry is approximately Euclidean. I will focus on the recovery or emergence of approximate isotropy (or rotational invariance) from the tile space, since this removes the main barrier for recovering of Euclidean geometry (for example, see Forrest's ([1995]) discussion on the anisotropy problem). Also, like other authors, the recovery of approximate Euclidean geometry is treated as approximate embeddability of physical quantities into Euclidean space (for example, see Chen [2021], Appendix A).\footnote{The difference is that the other work in the literature seeks to embed distances to Euclidean space as part of the fundamental physical quantities, while in this paper, the aim is to embed all the observable quantities of a given dynamics---albeit very rudimentary ones---to Euclidean space. The idea is that large-scale distances as empirical quantities supervene on these observables.}

	\section{Case One: Random Walk}
	
	In this section, I will present a toy example of random walks to illustrate how Euclidean geometry can emerge from the dynamics that does not presuppose it. 	 Random walks are usually studied in statistical physics, which is a branch of physics that studies how properties of macroscopic systems arise from stochastic microscopic motions. As a proof of concept, this example is chosen for its mathematical simplicity and intuitiveness as well as its conceptual rigor, even though it is far removed from the description of a physically realistic situation. 
	
	 Following Weyl, I will focus on a two-dimensional tile space  for simplicity.   Imagine  a tiny cat running from tile to tile. From each tile, the cat can move along four directions, characterized by two basis vectors $\pm e_1,\pm e_2$ (with $|e_i|$=1 for all $i$), to one of the neighbouring tiles for each time step. Suppose this is the only physical law in this simple world. The quantities we shall focus on are the probabilities of the cat showing up at certain tiles at certain times. I will show that the probability distribution over the tile space is approximately isotropic. That is, there is a probability-preserving embedding from the tile space into Euclidean space with an approximately isotropic probability distribution. More precisely, I will show:

		 \begin{quote}
		 	\textsc{The Isotropy Theorem}. For any starting position $x\in \mathbb{Z}^2$, and for any two tiles $y,z\in\mathbb{Z}^2$ such that $|y-x|\approx |z-x|$, the probabilities of the cat showing up in $y$ and $z$, if nonzero, are approximately the same after sufficiently long time.\footnote{
		 		The qualification of `non-zero' is imposed because for any number of steps, the probability of the cat showing up at any tile with the opposite parity is zero. For example, after any even number of steps, the probability of the cat at any tile represented by $(x,y)$ with $x+y=odd$ is zero.
		 		
		 		This result is similar to a central theorem in probability theory, the \textit{central limit theorem} (CLT), according to which the normalized sum of independent random variables (under certain conditions) tends towards a normal distribution (see van der Vaart [1998]). We can apply this theorem to our case of a random walk. CLT implies that as time tends to infinity, the distribution of the probability converges to that of the $n$-dimensional normal distribution centred around the starting position. More specifically, after $n$ steps, the probability of the cat showing up at a region around a particular tile with the size $O(\sqrt{n})$ is rotationally invariant as $n$ tends to infinity. The isotropy theorem  is stronger than this general result because it is about the approximate isotropy of the probability distribution on single tiles rather than large regions.} 
		 \end{quote}
	
\paragraph{Proof}		 
Let the starting position of the cat be the origin $(0,0)$.	It's easy to see that, after the first step, the probability of the cat showing up in each of the four neighbouring tiles is 1/4. Our goal is to calculate the probability for any tile after sufficiently many steps.  If the cat could not backtrack and only move in two directions, then such a probability would be easy to calculate. In that case, to reach $(x,y)$ after $n$ steps, the cat needs to take a total of $x$ steps in the `horizontal' direction and a total of $y$ steps in the `vertical' direction. Using basic combinatorics, the result would be $(1/2)^n\binom{n}{x}$, since each path the cat might take has a probability of $(1/2)^n$ and there are $\binom{n}{x}$ possible paths. The problem is trickier now that the cat can move back and forth along horizontal and vertical directions.\footnote{The obvious idea in the simple case would not work. For example, let the number of steps along the four directions be $a,b,c,d$ respectively. The probability of reaching $(x,y)$ after $n$ steps is equal to $(1/4)^n\binom{n}{a}\binom{n-a}{b}\binom{n-a-b}{c}$, with $a+b=x$ and $c+d=y$. But this probability cannot be calculated because there are too many unknown variables. } Fortunately, with a clever trick, we can obtain that the probability of reaching $(x,y)$ after $n$ steps, if nonzero (that is, when $x+y$ is of the same parity as $n$):
		\begin{align}
		(1/4)^n\binom{n}{(n+x+y)/2}\binom{n}{(n+x-y)/2}.
		\end{align}
		See Appendix A.1 for the detailed proof. 
		
Furthermore, it follows from a result in (Gallager [1968]) that	(1)  is approximately equal to the following when $x\ll n, y\ll n$ (see Appendix A.2 for the proof):\footnote{In other words, the result is conditioned on that enough time has lapsed relative to the locations of measurements. Thus, the observable isotropy in the random walk case is not entirely scale-independent: it depends on the relative scales between space and time. } 
		\begin{align}
	\frac{2e^{-(x^2+y^2)/n}}{\pi\sqrt{n^2-(x^2+y^2)}}.
		\end{align}
The result (2) only contains the Euclidean norm $x^2+y^2$ and is therefore rotationally invariant.\footnote{This is a recurring form in the cases that have approximate isotropy: terms other than the Euclidean norm or its powers disappear or are negligible (see also (27) in Appendix B.1 and (37) in B.3). } This can be straightforwardly generalized to any tile as the starting position (as it is essentially a matter of coordinate translation). This concludes the proof for the isotropy theorem. $\square$  

\begin{figure}[h]
	\centering
	\includegraphics[width=0.8\linewidth]{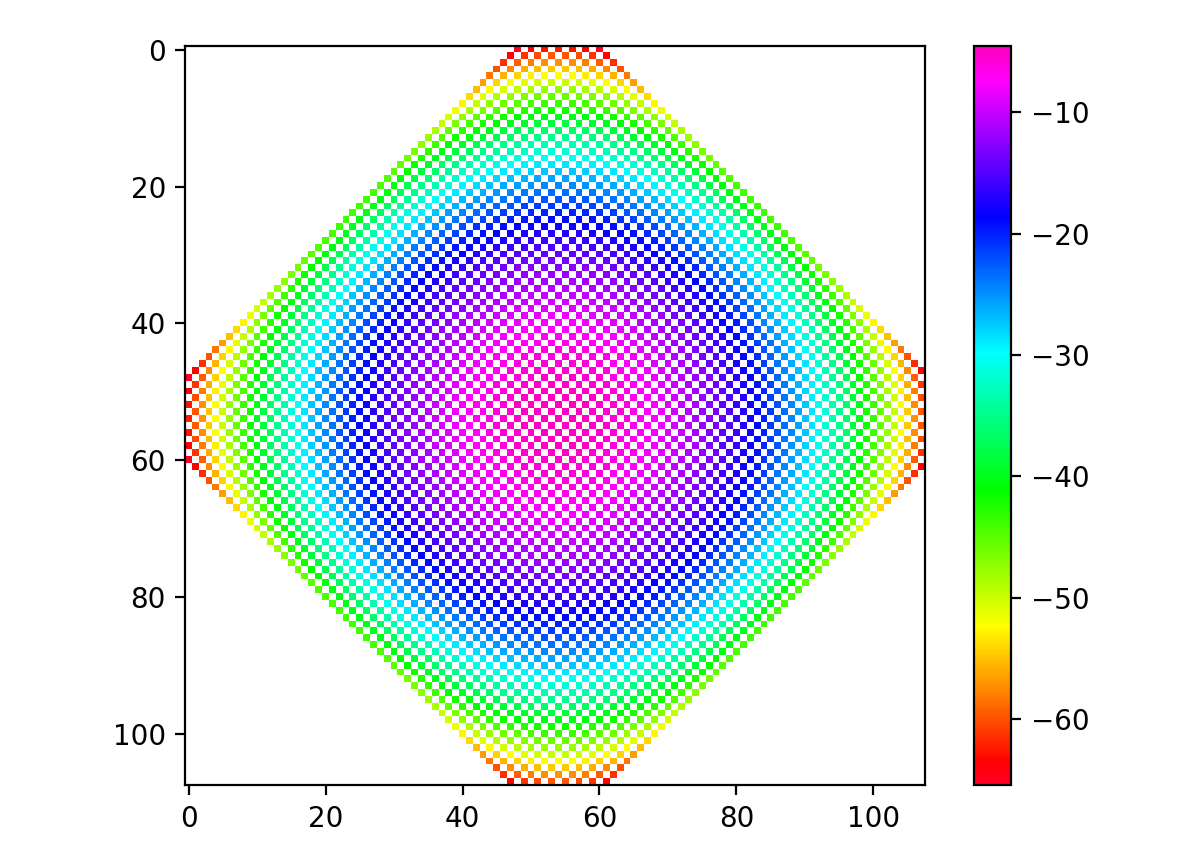}
	\caption{the probability distribution of the cat at $t=60$ on the tile space. Note that the colours correspond to the \textit{logarithm} of the probabilities in order to differentiate between small probabilities.}
	\label{fig:rwplots}
\end{figure}

Now that the proof is given, we can further compute and visualize the probability distribution of the cat at a time to provide an intuitive picture of what it is like. Figure 1 shows the probability distribution of the cat at $t=60$ as an example.\footnote{The codes that produce the figures in this paper are available upon request.} We can see that it is indeed approximately isotropic: the significant probabilities %represented by purple and deep blue colours 
form circle-like shapes. (Note that the white areas are tiles that the cat cannot reach at $t=60$, and we can see that all the tiles with the opposite parity of $t$ are indeed white.)

			It follows from the isotropy theorem that there is an embedding from the tile space into Euclidean space that approximately preserves the probability distribution of the cat after sufficiently long time (or more strictly speaking, from regions of the tile space of sizes suitably greater than one to regions of Euclidean space so that the effects of white tiles can average out). Assuming that such probabilities are our only observations, this means that our observations would be about the same as in Euclidean space. In this sense, the apparent Euclidean geometry emerges from the tile space under the dynamics of random walks.
			
			I would like to highlight some important features of this example.	First, in this example, there is no isotropic (or approximately isotropic) spacetime structure existing prior to dynamics. Indeed, I haven't defined a metric structure, and the topology is far from isotropic. It is also worth emphasizing that the Euclidean distance that occurs in the isotropy theorem is just a feature of our representation of the tile space, which allows us to  specify the desired embedding. Emphatically, it does not enter the dynamics nor plays a part in the proof for the theorem. Second, the dynamical law posited in this example is not contrived or overly complicated. Indeed, it is the simplest one studied in statistical field theory. The simplicity of laws is an important feature that distinguishes this case from the simulation of continuous mechanics on discrete pixels in computer programs. In such programs, the description of (say) a rotation either makes use of Euclidean geometry where the discrete pixels are embedded or gets very complicated. This complexity is one main reason why we typically do not consider reality as composed of pixels on which continuous motions are simulated. But we should reconsider that in light of the current example. 
			
			Also, note that the recovery of isotropy in this case does not critically involve macroscopic regions consisting of a vast number of tiles. Rather, the statistical correlations between individual tiles are already almost isotropic. This makes the case radically different from---for example---Bendegem's ([1987], [1995]) proposals, which crucially rely on macroscopic geometric entities for recovering Euclidean geometry (see Footnote 2).

			Of course, the random walk case is physically unrealistic: it does not depict the behavior of any fundamental particle or field; our space is not two dimensional; actual observations are vastly more complicated, to name a few. The physical situations most relevant to this case are those studied by statistical field theory, such as the transmission of heat (the transmission of heat would be isotropic even if the molecules involved moved along discrete tiles). But this is not a realistic interpretation of the current example, since the molecular movements occur at much larger scales than the fundamental unit of space.
			
			Nonetheless,  this case already constitutes serious evidence against the implicit assumption in the Weyl tile argument that the empirically observable distances are determined by counting the number of tiles. Call a dynamical law `isotropic' (or `rotationally invariant' or `has rotational symmetry') if its application to a system is not affected by any rotation of the system (mathematically, we can say that the application of the law to a system commutes with any rotation of the system).  The isotropy theorem shows that the simple law of random walks is isotropic with regards to a single cat (and it is not difficult to check that this is the case for any number of cats), despite the underlying space being discrete and not isotropic.  This suggests that it is also possible that our fundamental laws are  isotropic even if spacetime is discrete. Therefore, it is a serious problem for Weyl's argument to exclude this possibility. The symmetries of the dynamical laws, which determine observational symmetries, may not be among the symmetries of spacetime. 
			
			 I intend this case to do the conceptual heavy lifting for my objection to the Weyl tile argument due to its conceptual clarity and simplicity, but I will now turn to the case of quantum mechanics for a more physically relevant example.

\section{Case Two: Quantum Mechanics}

In this section I will show how observational rotational invariance can emerge from quantum mechanical systems in discrete space that does not presuppose any isotropic geometric structure.  The  sketched proof in the main text is intended to be accessible to philosophers with some conceptual familiarity with the formalism of quantum mechanics, with more mathematical details left to Appendix B.  Despite this case being more complicated and subtle than the random walk case, its relevance to fundamental physics can bolster the sentiment that it is really possible that our dynamical laws are written on discrete space (or spacetime) and are approximately isotropic for all we know.
	
For generality, I will consider an $n$-dimensional tile space. Let each tile be represented by a member of $\mathbb{Z}^n$. From each tile, there are $2n$ directions characterized by $n$ basis vectors $\pm e_1,\pm e_2,...,\pm e_n$ ($|e_i|=1$ for all $i$). I will show that the following claim is true:

\begin{quote}
	\textsc{The Quantum Isotropy Theorem}. For any quantum mechanical system with its initial position spread out in a sufficiently large region $A\subset\mathbb{Z}^n$, its time evolution is isotropic: the evolution of the system approximately commutes with any rotation of it.\footnote{ `Spread out' means that the position wavefunction of the system does not vary much in short distances.  Very roughly, a rotation of a system relative to position $a\in \mathbb{Z}^n$ as a wavefunction maps its values at every $x\in \mathbb{Z}^n$ to $y\in\mathbb{Z}^n$ that has similar Euclidean distance to $a$ as $x$.  A more rigorous, mathematical description of rotation can be found in Appendix B. 
}
\end{quote} 
As a special case, if the starting position $A$ is spread out and rotationally invariant, then the wavefunction will continue to be rotationally invariant: for any $y,z\in\mathbb{Z}^n$ that have approximately the same Euclidean distance to (the centre of) $A$, the amplitudes of the system at $y$ and $z$ at any time are approximately the same.\footnote{A centre of a region is a tile that minimizes the maximal Euclidean distance to other tiles in the region. As explained earlier, `Euclidean distance' refers only to the geometry of the Euclidean space where we embed the tile space, not to a primitive geometric structure of the tile space.} This implies that for any two regions that consist of tiles with similar Euclidean distances to the starting region, the probabilities of observing the system  in those regions are approximately the same. Assuming that such probabilities are our only observables, there is an embedding from the tile space to Euclidean space that preserves all structures and observations, as in the previous case. Notice that  the quantum isotropy theorem requires the initial position to be spread out, which is why this result is conceptually weaker than the isotropy theorem.\footnote{Another difference between this case and the first one, as we will see, is that the time is not discrete in this case. Unlike the first case, for any time $t>0$, the position wavefunction is nowhere zero. Heuristically, we can think of continuous time as composed of infinitely many infinitesimal durations, and therefore for any finite time, the `Schrödinger's cat' could show up anywhere.} But this does not necessarily mean it is less satisfactory, as it is possible that we can only observe or prepare a system spread out in a suitably large region.\footnote{A participant at the 2022 APA pacific division meeting asked whether this places a constraint on the initial condition of the universe and whether this is acceptable. While I find the question intriguing, the current case is only intended to be suggestive, since quantum mechanics is not fundamental. Thus I refrain from reading too much into the case.}

Before I go into the proof, I want to forestall a dismissive reaction. Some people (for example, from physics communities) may find	the quantum isotropy theorem obvious or even a stronger version of it that does not require the initial position to be spread out obviously true,  and therefore think there is no need for a demonstration.\footnote{Why might some people think so? I cannot hope to explain in a way that does justice to all those who have such beliefs, but here's a sketch of the reason why. The gist is that a momentum wavefunction is approximately isotropic for small momenta. Small momenta correspond to large distances. Thus, focusing only on large distances, the position wavefunction is approximately isotropic. As we will see, this reasoning has some semblance of my proof, but is invalid.}  However, this reaction is unjustified. To show this, I demonstrate in Appendix B.3 that a stronger claim (which some take to be true) is actually false: for any quantum mechanical system with initial position $x\in \mathbb{Z}^n$, its wavefunction will not evolve to be approximately isotropic for any significant period of time. To the aforementioned readers, this negative result may be the most interesting one in this paper. Regardless, it is useful to have a negative claim to contrast the positive theorems of the paper with.

\paragraph{Proof}
To prove  the quantum isotropy theorem, we shall start from the Schrödinger equation, which governs the evolution of quantum mechanical systems (setting the Planck constant $\hbar$ to one):
\begin{align}
	i\frac{d}{dt}\Psi(t)=\hat{H}\Psi(t),
\end{align}
  where $\Psi(t)$ is the position wavefunction of the system at $t$, which assigns a complex-valued amplitude to each spatial point, and $\hat{H}$ is the Hamiltonian operator on the wavefunction (which indicates the total energy of the system). In order to apply the Schrödinger equation to discrete space, we need to formulate the discrete version of the equation. Since I will exclusively consider the discrete case, I will use the same notation without risking ambiguity. First, the discrete version of the wavefunction $\Psi(t)$ for any given $t$ can be considered a complex-valued function over the tile space: $$\Psi(t): \mathbb{Z}^n\to\mathbb{C}.$$  For the right side of (3), we need to discretize the Hamiltonian. If we set the mass  of the system under consideration to one and ignore its potential energy, the Hamiltonian is equal to its kinetic energy $\displaystyle-\frac{1}{2}\sum_i(\frac{\partial}{\partial x^i})^2$. A natural discrete definition of $\frac{\partial}{\partial x^i}$ would be the difference of the value of a given function between neighbouring tiles along a certain direction. That is: 
\begin{align}
	\frac{\partial}{\partial x^i}\Psi (t,x) &=\Psi(t, x+e_i)-\Psi(t,x) 	\textrm{, or} \\
	& =\Psi(t,x)-\Psi(t, x-e_i).
\end{align}  
Then, the discrete version of the Hamiltonian is this (concerning only its kinetic energy part):\footnote{This is obtained by applying $\frac{\partial}{\partial x^i}$ twice in opposite directions (that is, (4) and (5) respectively). If we only apply (4) twice, then we would have $$\hat{H}\Psi(t,x)=-\frac{1}{2}\sum_i(\Psi(t, x+2e_i)-2\Psi(t,x+e_i)+\Psi(t,x)),$$ which is also a legitimate choice, but makes calculation more complicated.} 
\begin{align}
	\hat{H}\Psi(t,x)=-\frac{1}{2}\sum_i(\Psi(t, x+e_i)+\Psi(t,x-e_i)-2\Psi(t,x)).
\end{align}
$\Psi(t)$ is a function in the position space, and it is useful to transform it into one in the momentum space, where we can prove its rotational invariance more easily. The inverse Fourier series of $\Psi(t)$ is its momentum space counterpart $\tilde{\Psi}(t): \mathbb{R}^n\to \mathbb{C}$:\footnote{$\Psi(t)$ is the \textit{Fourier series} of its momentum counterpart $\tilde{\Psi}(t)$ because the domain of $\Psi(t)$ is discrete. In (7), we can see that the momentum space wavefunction is the discrete sum of a series.}
\begin{align}
	\tilde{\Psi}(t,p)= \sum_{x\in\mathbb{Z}^n} e^{-2\pi i p x} \Psi (t,x),
\end{align}
where `$ p x$' is an abbreviation for the inner product of n-vectors $p$ and $x$. Note that this momentum wavefunction is periodical. That is, $\tilde{\Psi}(t,p)=\tilde{\Psi}(t,p+e_i)$.\footnote{Here's the derivation (omitting $t$ for brevity):
	$$\tilde{\Psi}(p+e_i)= \sum_{x\in\mathbb{Z}^n} e^{-2\pi i (p+e_i)x} \Psi (x)=\sum_{x\in\mathbb{Z}^n} e^{-2\pi i px}e^{-2\pi i x_i} \Psi (x)= \sum_{x\in\mathbb{Z}^n} e^{-2\pi i px} \Psi (x)= \tilde{\Psi}(p).$$}
Therefore, we can consider $\tilde{\Psi}(t)$ as a complex-valued function defined on the quotient space $\mathbb{R}^n/\mathbb{Z}^n$, which means that we can `collapse' all the $\mathbb{R}^n$ points with integer distances away into one point. This quotient space can be represented by the unit square around the origin: $B=[-1/2,1/2]\times [-1/2,1/2]$ (`$B$' stands for `Brillouin zone'). Then we have $\tilde{\Psi}(t):B\to\mathbb{C}$. We can transform it back to the position wavefunction in the following way:
\begin{align}
	\Psi (t,x)=\int_{p\in B} e^{2\pi i px}\tilde{\Psi}(t,p)dp.
\end{align}
It follows that the time evolution of $\Psi$ is approximately isotropic if the evolution of $\tilde{\Psi}$ is approximately isotropic (see Appendix B.2 for the proof). Moreover, we can show that, assuming that the initial position wavefunction $\Psi(0)$ is sufficiently spread out, the time evolution of  $\tilde{\Psi}$ is indeed approximately isotropic (see Appendix B.1 for the proof). The quantum isotropy theorem follows. $\square$

 The proof can be corroborated by a simulation of the time evolution of the position wavefunction, which offers a more intuitive picture. In Figure 2, we start with an approximately isotropic wavefunction that is spread over tiles represented by $(x,y)$ with $x^2+y^2<10^2$,  and compute its evolution over time. The wavefunction is plotted at $t=0, t=30,$ and $t=300$ respectively. 
   \begin{figure}[h]
 	\centering
 	\includegraphics[width=1\linewidth]{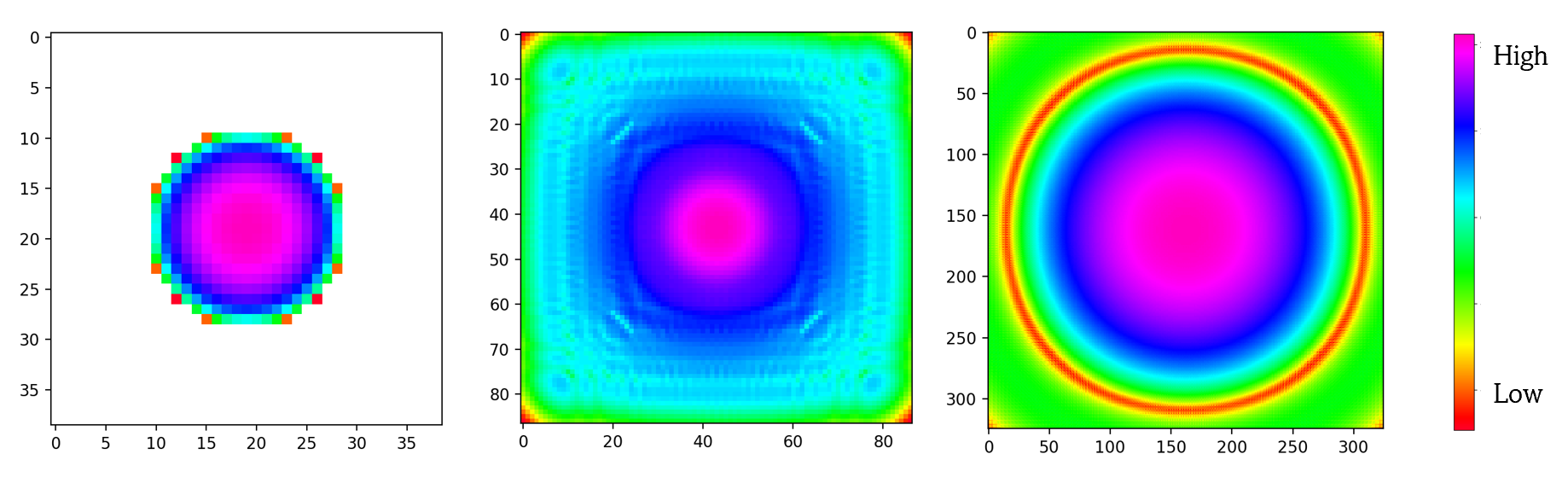}
 	\caption{The evolution of a wavefunction with initial radius of 10. All significant parts of the wavefunction are plotted. 	(Left)  $t=0$; (Middle) $t=30$; (Right) $t=300$. Note that the scales are different between the plots, since the wavefunction is more spread out as time passes; also, the values represented by the same colours are different between the plots, since the amplitudes generally get much lower as the wavefunction spreads thin.} 
 	\label{fig:qmplots}
 \end{figure}
 We can observe that in all these times, the wavefunction is indeed approximately isotropic: the significant parts of the wavefunction %represented by purple and dark blue pixels 
 at these times have circle-like shapes. But we can also see that the approximate isotropy gets more perfect as time passes. In early times such as $t=30$, we can still observe vertical and horizontal stripes,
%in blue and green
   which are interference patterns. These almost disappear in later times such as $t=300$: for example, the destructive interference stripes 
   %in red and yellow
    are almost perfect circles. In contrast, in Figure 4 in Appendix B.3, the evolution of a wavefunction with its initial position at a single tile looks very different and is never isotropic---its interference stripes are always perpendicular. (This stark contrast might surprise some readers given that the spread-out wavefunction simply `sums over' copies of the other.)

Does the isotropic time evolution of quantum mechanical systems mean the emergence of Euclidean geometry at the empirical level? Strictly speaking, observable Euclidean geometry is not yet recovered since the apparent distances are not even defined. But this is not a problem in principle: assuming quantum mechanics were fundamental, all matters, including length-measuring devices, are reducible to quantum mechanical systems (although a more realistic discussion would need to appeal to quantum field theory or a more fundamental future theory). Thus, ensuring that the observable quantities in the fundamental theory are isotropic also ensures that the apparent distances are Euclidean.

While this case has more physical relevance than the random walk case, this is still a very simplistic example. For one thing, quantum mechanics is not a fundamental theory due to its conflict with relativity (in the current example, the temporal dimension is still pre-relativistically separate from the spatial dimensions and is not discretized).  Quantum field theory is more fundamental, and there is indeed a sentiment in the community that a large-scale isotropy will emerge from fundamental laws defined over discrete spacetime.\footnote{This is a common hope in the community of lattice quantum field theory (LQFT) (for example, see Montvay and Munster [1994]). Note that the lattices involved in LQFT are not necessarily intended to be read realistically, but as a computational device.} But to convert such a sentiment into rigorous frameworks, theorems, and proofs, there is still a lot to be done and nothing certain can be said at this stage. Furthermore, even quantum field theory is not sufficiently realistic for our purposes, since it does not address the problem of quantum gravity (that is, incorporating the gravitational field into the framework of quantum field theory). A successful handling of quantum gravity is important to a realistic story concerning the fundamental structure of spacetime.\footnote{It might be worth mentioning that in (Chen [forthcoming]), I argue that we do not have a realistic understanding of spacetime even at the experimentally accessible level in the absence of a successful theory of quantum gravity. 
	
	The research programs of quantum gravity that appeal to discrete spacetime are conceptually similar to the project of this paper, only with vastly more complexity: they involve developing physics on discrete spacetime that yields general relativity as its limit. See for example (Hamber [2009]).} 

As a further observation, we note that in both this case and the random walk case, scale plays a part in the recovery of approximate isotropy. In the random walk case, the observable isotropy is conditioned on the relative scales of the time and the location of measurements (see Footnote 9). Here, the initial wavefunction is required to be sufficiently spread out for it to evolve isotropically. It is helpful to compare this with the geometricist solutions to Weyl's argument, where approximate Euclidean geometry emerges only at a sufficiently large scale (van Bendegem [1987], [1995], Forrest [1995]). While some sort of scale-dependence occurs in all the approaches, it is worth noting two differences between the current dynamical approach and the rest. First, unlike the geometricist solutions, the scale involved in our cases is not a fundamentally metric-theoretic notion, since there is no metric structure. Second, our approach dispenses with a  primitive distinction between the macroscopic scale and the microscopic scale postulated in the geometricist solutions (see Footnote 2). It is rather derived as a consequence of the dynamical laws.

		\section{Two Dogmas of Geometricism }
		
		So far I have laid out the main results and my objection to Weyl's argument, but it is worth expounding on where the argument goes wrong. In particular, I will criticize two (interrelated) assumptions underlying it: (1)  large-scale or observable physical geometry is determined by fundamental spacetime structures independent of dynamical laws;\footnote{Note that by `large-scale' or `observable' geometry, I do not mean measurements we perform to detect geometrical features. Clearly, such measurements are not independent of the mechanics of the measurement devices, and geometricists would not deny this. Rather, I mean the large-scale geometry that is tracked by such measurements.} (2) some geometric structures including the metric structure are ontologically and explanatorily prior to dynamics and must be presumed by the latter.  These are what I shall call the `two dogmas' of geometricism, the view that geometry is more fundamental than dynamics, with the opposite view called `dynamicism'.\footnote{Of course, I do not intend the two dogmas to reflect all variants of geometricism. For example, some geometricists (see Maudlin [1988], [2012]) consider the metric structure as an essential feature of spacetime, while others (such as Earman and Norton [1987], Norton [2008]) think that spacetime is represented by metrically amorphous manifolds and the metric is a matter field in spacetime. But I do believe that the two dogmas are commonly held by geometricists and philosophers at large. This is understandable. Classical dynamical laws often presuppose geometrical notions. For example, the law of inertia in Newtonian mechanics says that a free system moves along a straight line, where straightness is a geometrical notion. In relativistic theories, an analogous law that says that a free-falling system moves along a time-like geodesic. Thus, it is natural to think that spacetime has a metric structure that dictates how matter in spacetime behaves. 
			} (The debate between the two positions usually proceeds in the case of continuous spacetime, but I think the discrete case examined in this paper can be more helpful in clarifying how dynamics can be more fundamental than geometry, and how we can have interesting physics without any fundamental metric structure.)

		First, the Weyl tile argument is wrong about how large-scale distances emerge from the fundamental structures. As pointed out in Section 1, the argument relies on 	the counting account, which assumes that distances and in general the physical geometry  can be determined independently from dynamical laws. In the two examples discussed in the paper, I have shown that dynamics play a crucial role in determining physical geometry. The observable isotropy is determined by the dynamical laws that govern the movement of the tiny cat in the random walk case and the evolution of the wavefunction in the case of quantum mechanics. In general, since we observe physical geometry with various measuring devices like rigid rods, it is natural to expect that the measurement we get is partly determined by how those devices work. (This is a hotly debated claim in the debate between geometricism and dynamicism; see for example Brown [2005], Maudlin [2012], Norton [2008], Menon [2019].)

Note that in the two cases, even if we posit a metric structure according to the counting account, or in any other ways, it would have no empirical consequences, since it does not play any role in the dynamical laws. So we should not posit such a structure, given that we should not posit structures with no empirical consequences.

Second, it is a mistake to assume that geometry must be presumed by dynamics. Hopefully this is already clear from the previous discussion, but is still worth emphasizing. The difference between this dogma of geometricism and the first one is this: one may grant that the apparent geometry arises from dynamical laws but still insist that some fundamental geometric structure must be presumed by dynamics. For example, in the famous Poincaré  disk scenario (Poincaré  1912[2018]), the geometry appears hyperbolic to the residents of the disk because of the dynamics (there is a universal force that shrinks rigid rods and bends light beams). But we know by stipulation that such dynamical laws are still defined over Euclidean geometry, which exists fundamentally. Thus the fact that dynamical laws play a role in determining apparent geometry does not necessarily mean there is no underlying `real' geometry. It is a common assumption that dynamical laws need to be `written on' spacetime geometry (Earman [1989], 46). But the two cases considered have demonstrated that the fundamental dynamical laws do not need a prior metric structure.\footnote{In his commentary, David John Baker asked if there is still a metric concerning the distinction between space and time. This is an interesting question, but in this paper I only focus on discrete space and the spatial metric. 
	
} The discrete version of the Schrödinger equation only requires the topological structure and the derived `differential' structure.

One may object that the topological structure or just the set of tiles itself is still a geometrical structure, and therefore I haven't refuted the second dogma. Indeed, Norton ([2008]) objected to dynamicism by arguing that spacetime coincidence is not derivable from dynamical laws but must be presupposed.  Fair enough---in this paper, I only intend to show that no metric structure needs to be presupposed by dynamics. To get rid of spacetime altogether, we need a very different framework such as algebraicism rather than the standard point-set-theoretic framework, which I will not delve into here (for example, see Geroch [1972], Connes [2013], Menon [2019], Chen and Fritz [2021]).\footnote{These authors did not discuss the discrete case explicitly. But the discrete case can be a special case of the formalism we proposed in (Chen and Fritz [2021]). I shall discuss this elsewhere, because algebraicism is a topic all by itself.}

	\section{Conclusion}
			
	The Weyl tile argument against discrete space implicitly assumes that the symmetries of geometry help determine symmetries of matter systems: since the geometry of the tile space is not isotropic, the physical laws and observables would also lack isotropy.  However,  I have shown that the observable physical states in the random walk case and quantum mechanics are decoupled from the tile geometry. Even though the tile space is radically non-Euclidean in the sense that the diagonal of a square contains twice as many steps as the side, the observable geometry can still be approximately Euclidean.

		\newpage
		
\appendix

\section{Random Walk}
			\newtheorem{random1}[subsection]{Theorem}
			\begin{random1}
					The probability of reaching $(x,y)$ from $(0,0)$ after $n$ steps is equal to zero (when $x+y$ has the opposite parity of $n$) or
					\begin{align}
				(1/4)^n\binom{n}{(n+x+y)/2}\binom{n}{(n+x-y)/2}.
			\end{align}
			\end{random1}

	\paragraph{Proof}
	 It should be obvious that the probability of reaching $(x,y)$ from $(0,0)$ after $n$ steps is equal to zero when $x+y$ has the opposite parity of $n$, so in what follows, I will assume $x+y$ has the same parity as $n$. We imagine that in every step the cat takes, the cat actually takes two half-steps along the diagonal directions. We assume that there are four possible ways the cat can move in one step from $(0,0)$: $(1/2,1/2)+(1/2,-1/2)$, $(1/2,1/2)+(-1/2,1/2)$, $(-1/2,-1/2)+(1/2,-1/2)$, $(-1/2,-1/2)+(-1/2,1/2)$. This corresponds to the four possible ways to move in the original situation $(\pm1,\pm1)$. We can consider the original situation and the imagined situation as two ways of representing the same physical situation. As an analogy, in chess, a knight's move can be equivalently considered as one L-shaped step or as consisting of first moving one row (or file) and then moving two files (or rows). Let's call the number of steps that the cat takes in the aforementioned four possible ways respectively $ac,ad,bc,bd$  (Figure 3).  
		\begin{figure}[h]
			\centering
			\includegraphics[scale=0.8]{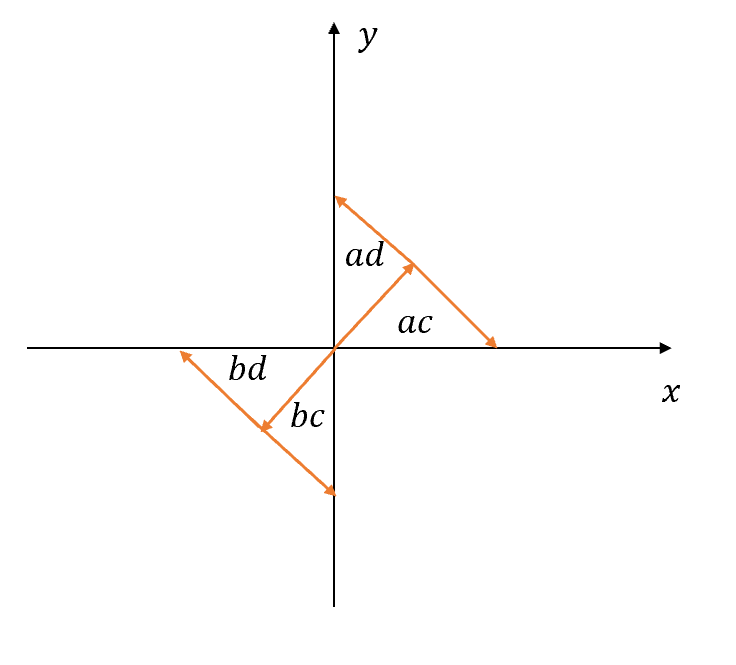}
			\caption{}

		\end{figure}
		Because the number of steps involving $a$ (namely $ac+ad$) and the number of steps involving $c$ (namely $ac+bc$) are independent, the probability of reaching $(x,y)$ after $n$ steps is equal to
		$$(1/2)^n\binom{n}{ac+ad}\cdot(1/2)^n\binom{n}{ac+bc}.$$
		with $ac+ad+bc+bd=n$ and $ac-bd=x, ad-bc=y$. It follows that the probability of reaching $(x,y)$ after $n$ steps is equal to
		$$(1/4)^n\binom{n}{(n+x+y)/2}\binom{n}{(n+x-y)/2}.$$

		\newtheorem{random2}[subsection]{Theorem}
		\begin{random2}
		 Assuming $x\ll n$ and $y\ll n$, 	(9) is approximately equal to 
	\begin{align}
\frac{2e^{-(x^2+y^2)/n}}{\pi\sqrt{n^2-(x^2+y^2)}}.
	\end{align}
	\end{random2}
		
		\paragraph{Proof}
	Assuming both $n$ and $n-k$ are very large, we have the following theorem (Gallager [1968], p.530): 
		\begin{align}
			\binom{n}{k}\to \sqrt{\frac{n}{2\pi k(n-k)}} e^{nh(k/n)}.
		\end{align}
		where $k\in [1,n-1]\cap\mathbb{Z}$ and $h=-x\ln x-(1-x)\ln (1-x)$. It is worth noting that (11), which is directly derived from Stirling's formula, is not based on any consideration from physical geometry, so we are not at risk of smuggling Euclidean geometry into our proof. Plugging (9) into (11), the result would be of the form $\alpha e^\beta$, where
		\begin{align}
			\alpha=(1/4)^n \sqrt{\frac{4n^2}{\pi^2 (n^4-n^2(x^2+y^2)+(x^2-y^2)^2)}}.
		\end{align}

	Since $x\ll n$ and $y\ll n$, we have $(x^2-y^2)^2\ll n^2$, and thus (12) is approximately equal to 
	\begin{align}
		(1/4)^n \sqrt{\frac{4n^2}{\pi^2 n^4-n^2(x^2+y^2)}}=(1/4)^n\frac{2}{\pi\sqrt{n^2-(x^2+y^2)}}.
	\end{align}
		Note that this result involves only the Euclidean norm $x^2+y^2$, which is rotationally invariant. This is the first approximation involved in the proof.
		
		The exponent $\beta$ is $nh(\frac{n+x+y}{2n})+nh(\frac{n+x-y}{2n})$. Let $u=x/2n$ and $v=y/2n$. Then:
		\begin{align}
		\beta=nh(\frac{1}{2}+u+v)+nh(\frac{1}{2}+u-v).
		\end{align}
	 Since $u+v$ and $u-v$ are close to zero, we can approximate $h(\frac{1}{2}+u+v)$ and $h(\frac{1}{2}+u-v)$ by expanding $h$ around $1/2$ up to the second order (let $p\in \mathbb{R}$ be small):
	 \begin{align}
	 	h(\frac{1}{2}+p)\approx h(\frac{1}{2})+h'(\frac{1}{2})p+1/2h''(\frac{1}{2})p^2=ln2-2p^2,
	 \end{align}
		since
		\begin{align*}
			h(\frac{1}{2})&=-1/2\ln 1/2-1/2\ln1/2 =\ln 2,\\
			h'(\frac{1}{2})&=-\ln 1/2+\ln 1/2=0,\\
			h''(\frac{1}{2})&=-2-2=-4.
		\end{align*}
		This is the second and last approximation involved in the proof. Plugging them into (14), we obtain:
		\begin{align}
		\beta &\approx	n (\ln 2 -2(u+v)^2) + n(\ln 2 -2(u-v)^2)\\
		&=n\ln 4-4n(u^2+v^2)=n\ln 4-\frac{x^2+y^2}{n}.
		\end{align}
Then, the exponential $e^\beta$ is equal to $e^{n\ln 4-\frac{x^2+y^2}{n}}=4^n e^{-(x^2+y^2)/n}$. Together with (13), we obtain that $\alpha e^\beta$ is equal to (10). $\square$

	\section{Quantum Mechanics}
	
	$\Psi(t): \mathbb{Z}^n\to \mathbb{C}$ is the wavefunction of a certain quantum mechanical system at $t$ in discrete space, with 	$\Psi(0)$ suitably spread out. $\tilde{\Psi}(t):\mathbb{R}^n\to \mathbb{C}$ is its inverse Fourier series in the momentum space defined by $\tilde{\Psi}(t,p)= \sum_{x\in\mathbb{Z}^n} e^{-2\pi i p x} \Psi (t,x)$. In this appendix, I will call tiles `lattice points' which is more standard in this context.

					\newtheorem{qm2}[subsection]{Theorem}
					
					\begin{qm2}
						Let $A$ be any rotation matrix in the momentum space. Assuming that the initial position space wavefunction $\Psi(0)$ is sufficiently spread out, we can show that the time evolution of $\tilde{\Psi}$ commutes with $A$. That is, $e^{-i\tilde{H}t}\tilde{\Psi}(0,Ap)\approx \tilde{\Psi}(t,Ap)$, where $e^{-i\tilde{H}t}$ is the operator that represents the time evolution up to $t$.  
					\end{qm2}
	
\paragraph{Proof}

By applying the Fourier series transform to $\hat{H}\Psi(t,x)$, we can get the momentum space Hamiltonian $\tilde{H}$  satisfying the Schrödinger equation $i\frac{d}{dt}\tilde{\Psi}(t)=\tilde{H}\tilde{\Psi}(t)$. We first observe that $\Psi(t, x+e_i)$ can be transformed to $e^{2\pi i p_i}\tilde{\Psi}(t,p)$ through the following steps (where $p_i=p\cdot e_i$, namely the $i$-th component of $p$): 
	\begin{align}
	\Psi(t, x+e_i) &\Rightarrow  \sum_{x\in\mathbb{Z}^n} e^{-2\pi i p x} \Psi (t,x+e_i) \\
	&= \sum_{x\in\mathbb{Z}^n} e^{-2\pi i p (x-e_i)} \Psi (t,x) \\
	&=  \sum_{x\in\mathbb{Z}^n} e^{-2\pi i p x}e^{2\pi i p e_i} \Psi (t,x)\\
	&= e^{2\pi i p_i} \tilde{\Psi}(t,p).
\end{align}
Similarly, 	$\Psi(t, x-e_i)$ can be transformed to $e^{-2\pi i p_i}\tilde{\Psi}(t,p)$.  Thus, from $\hat{H}\Psi(t,x)=-\frac{1}{2}\sum_i(\Psi(t, x+e_i)+\Psi(t,x-e_i)-2\Psi(t,x))$ we can obtain the following:
\begin{align}
	\tilde{H}\tilde{\Psi}(t,p)=-\frac{1}{2}\sum_i(e^{2\pi i p_i}+e^{-2\pi i p_i}-2)\tilde{\Psi}(t,p).
\end{align}
Let's abbreviate the expression `$-\frac{1}{2}\sum_i(e^{2\pi i p_i}+e^{-2\pi i p_i}-2)$' as `$\mathbf{Cow}(p)$'. Then, $\tilde{H}\tilde{\Psi}(t,p)=\mathbf{Cow}(p) \tilde{\Psi}(t,p)$. Now we can solve the Schrödinger equation in the momentum space:
\begin{align}
	\tilde{\Psi} (t,p)&=e^{-i\tilde{H}t}\tilde{\Psi}(0, p)=e^{-i\mathbf{Cow}(p)t}\tilde{\Psi}(0,p).
\end{align}
By assumption, the quantum mechanical system under consideration starts with a wavefunction sufficiently spread out around the origin at $t=0$. That is, $\Psi (0,x)$ does not vary much over small distances. Given that $\tilde{\Psi}(0,p)= \sum_{x\in\mathbb{Z}^n} e^{-2\pi i p x}  \Psi (0,x) $, when $p$ is large,  $e^{-2\pi i p x} \Psi (0,x)$ tend to cancel off over small variation of $x$, so the results add up small (`destructive interference'). When $p$ is very small ($p\ll 1$), then there is no such destructive interference, and therefore the sum is much more significant.   So $\tilde{\Psi}(0,p)$ is negligibly small for large values of $p$ and only nonnegligible for small values of $p$.

We can show that $\mathbf{Cow}(p)$ is approximately rotationally invariant (a spherical cow) when $p$ is sufficiently small:
\begin{align}
	\mathbf{Cow}(p) &=-\frac{1}{2}\sum_i(e^{-2\pi ip_i}+e^{2\pi ip_i}-2)\\
	&= -\sum_i(cos(2\pi p_i)-1) \\
	&=  -\sum_i(1-2\pi^2p_i^2-1+O(p_i^4))\\
	&=2\pi^2\sum_ip_i^2 - O(p^4).
\end{align} 
Here $\sum_ip_i^2$ is the square of the Euclidean length of $p$ in the momentum space and is rotationally invariant. $O(p^4)$ is much smaller and can be ignored if $|p|$ is sufficiently small. 

%Now if we look at the whole integral, we can see that large values of $|p|$ does not contribute much to the integral because $\tilde{\Psi}(0,p)$ is negligibly small for large $|p|$. So only small values of $|p|$ make main contribution to the integral. Since we can ignore $O(p^4)$ when $|p|$ is sufficiently small, then the whole integral is approximately invariant under any rotation of $p$ in $\mathbf{Cow}$. 

Then, we obtain the desired result:
$$\tilde{\Psi}(t,Ap)= e^{-i\mathbf{Cow}(Ap)t}\tilde{\Psi}(0,Ap)\approx e^{-i\mathbf{Cow}(p)t}\tilde{\Psi}(0,Ap)=e^{-i\tilde{H}t}\tilde{\Psi}(0,Ap)\textrm{. } \square$$

		\newtheorem{qm1}[subsection]{Theorem}
\begin{qm1}
	If  the time evolution of $\tilde{\Psi}$ is approximately isotropic, then the evolution of $\Psi$ is also approximately isotropic: for any $t$ and any rotation $A$ acting on $\Psi$, $e^{-iHt}\Psi(0,Ax)\approx \Psi(t,Ax)$. (A rotation on discrete space can be considered an equivalence class of rotations on continuous space that map each point to places near the same lattice point.)
\end{qm1}
\paragraph{Proof}

Let $A$ be any rotation matrix on $\mathbb{R}^n$. We first extend $\Psi(t)$ to a continuous function $\Psi^+(t)$ over $\mathbb{R}^n$ with $	\Psi^+(t,x)=\int_{p\in B} e^{2\pi i px}\tilde{\Psi}(t,p)dp$. Then we can show that if  $e^{-i\tilde{H}t}\tilde{\Psi}(0,Ap)\approx \tilde{\Psi}(t,Ap)$, then $e^{-i\hat{H}t}\Psi^+(0,Ax)\approx \Psi^+(t,Ax)$.

We have: 
\begin{align}
	\Psi^+(t,Ax)&=\int_{p\in B} e^{2\pi i pAx}\tilde{\Psi}(t,p)dp \\
	&=\int_{p\in AB} e^{2\pi i ApAx}\tilde{\Psi}(t,Ap)dp \textrm{ (substitute $dp$ by $dA^{-1}p$)}\\
	&= \int_{p\in B} e^{2\pi i px}\tilde{\Psi}(t,Ap)dp \textrm{ ($ApAx=px$)}\\
	&\approx \int_{p\in B} e^{2\pi i px} e^{-i\tilde{H}t}\tilde{\Psi}(0,Ap)dp \textrm{ ($e^{-i\tilde{H}t}\tilde{\Psi}(0,Ap)\approx \tilde{\Psi}(t,Ap)$)}\\
	&= \int_{p\in B} e^{2\pi i pAx} e^{-i\tilde{H}t}\tilde{\Psi}(0,p)dp \textrm{ (like (29), (30))}\\
	&=e^{-i\hat{H}t}\Psi^+ (0,Ax).
\end{align}
In (29), we substitute the integration variable $p$ by $A^{-1}p$, so we integrate over the rotated Brillouin zone $AB$ rather than $B$, and all the occurrences of $p$ in (29) are replaced by $Ap$. To derive (30), we note that integrating over $AB$ is about the same as integrating over $B$ because only small values of $p$ around the origin make main contributions to the integral  (see B.1), which are all included in $AB$. Moreover, $ApAx=px$ because any rotation matrix is orthogonal, which means that it preserves the inner product of two vectors. From (32) to (33) we apply the Fourier transform which commutes with the time evolution.

The result (33) means that the time evolution of the extended position wavefunction (over $\mathbb{R}^n$) is  approximately isotropic. Of course, when $x$ is a lattice point, $Ax$ may not be. But $Ax$ and its nearest lattice points have very similar $\Psi$-values because changing $|x|$ by one in the above equations makes very little difference to the result (because the initial wavefunction is spread out by assumption, and only large $x$s contribute significantly to the integral above). Therefore, we can conclude that the time evolution of the original position wavefunction is also approximately isotropic. Theorem B.2 follows. $\square$

		\newtheorem{qm3}[subsection]{Theorem}
\begin{qm3}
	
	For any quantum mechanical system with initial position $x\in \mathbb{Z}_n$, its position wavefunction will not evolve to be approximately isotropic for any significant period of time, that is, it is never lastingly the case that for any $y,z\in\mathbb{Z}_n$ with $|y-x|\approx|z-x|$, the amplitudes of the wavefunction at $y$ and $z$ are approximately the same.
	\end{qm3}
Note that it is possible for the wavefunction to be approximately isotropic for a split-second (including the initial moment), but this has no significance for our observations---thus the qualification `any significant period of time'.

\paragraph{Proof.} The main difference between what this theorem falsifies and the quantum isotropy theorem is that the initial position in this theorem is a single lattice point rather than being spread out. Thus, all the reasoning for the quantum isotropy theorem that does not rely on the spread-out-ness of the initial position also applies here. What does not apply is the claim  that the momentum wavefunction $\tilde{\Psi}(t)$ is approximately isotropic (Theorem B.2). We can then translate this negative result into the position space and arrive at the conclusion that the position wavefunction is also not approximately isotropic. 

Recall that an important step for proving the approximate rotational invariance of $\tilde{\Psi}(t)$ is that $\tilde{\Psi}(0,p)$ is negligibly small for large values of $p$ and only nonnegligible for small values of $p$. But this is not the case if $\Psi(0)$ is not spread out. Suppose $\Psi(0)$ is 1 at point zero and 0 elsewhere. In this case, we have $\tilde{\Psi}(0,p)=1$. Then:

  \begin{align}
	\Psi (t,x) &=\int_{p\in B} e^{2\pi i px}\tilde{\Psi}(t,p)dp\\
	&=\int_{p\in B} e^{2\pi i px}e^{-i\mathbf{Cow}(p)t}\tilde{\Psi}(0,p)dp\\
	&=\int_{p\in B}e^{2\pi i px}e^{-i\mathbf{Cow}(p)t}dp\\
	&=\int_{p\in B}e^{2\pi i px}e^{-i(2\pi^2\sum_ip_i^2 - O(p^4))t}dp.
\end{align}
Here, the influence of $O(p^4)$ is nonnegligible because the contribution of a neighbourhood of large $p$ (that is, $|p|$ is close to 1/2) to the integral is just as significant as a neighbourhood of small $p$ (that is, $|p|$ is close to zero---far smaller than 1/2) as far as the coefficient $e^{2\pi i px}$ is concerned, so the `error' caused by $O(p^4)$ for large $p$ cannot be suppressed by the coefficient. But how large is the deviation? For this question, it is helpful to provide some numerical analysis.

We can evaluate $\Psi(t)$ through the modified Bessel function $I(t)$ by the following equation ($n$ is the dimension of space) (see Lemma B.4):
\begin{align}
	\Psi (t,x)=e^{-nit}\prod_{j\le n}  I_{x_j}(it).
\end{align} 
For brevity, let's consider two-dimensional space, in which case we have
\begin{align}
\Psi(t, x,y) = e^{-2it} I_x(it) I_y(it),
\end{align} 
 where $x,y\in \mathbb{Z}$ are the two spatial coordinates. 
 
 Then we can plot the modified Bessel function at a time $t$ over regions where the amplitudes are significant. (Note that the `witnesses' to the isotropy violation are always regions where the amplitudes are significant, for if we measure the amplitudes far out in space where the values are negligibly low then \textit{ipso facto} their differences are negligible.) As we can see in Figure 4, the wavefunctions are not approximately rotationally invariant. In each figure, there are obvious horizontal and vertical stripes representing different amplitudes from those of nearby points. These are interference patterns, which are responsible for the violation of isotropy, and such patterns will occur at all later times (we can contrast this with Figure 2, where interference patterns are never as obvious and get rounder over time). 

\begin{figure}[h]
	\centering
	\includegraphics[width=1\linewidth]{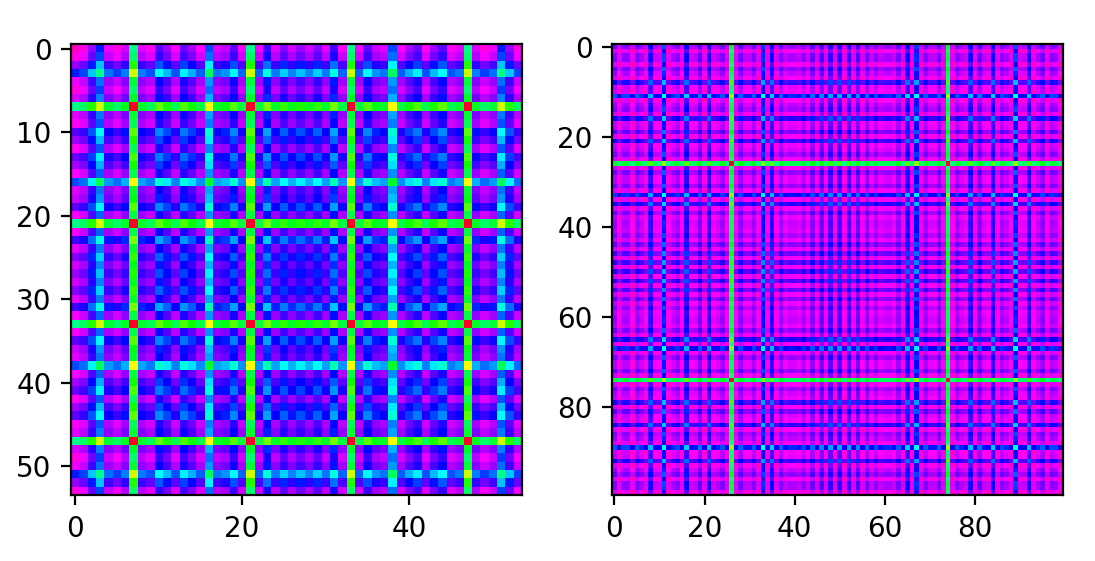}
	\caption{ The time evolution of a quantum mechanical system with its initial position at a single lattice point. (Left) $t=30$; (Right) $t=100$. As before, the scale of each plot is adjusted so that we focus on the significant part of the wavefunction. The plait shirt pattern remains for later $t$s, with only the colours and frequencies of the stripes changing.}
	\label{fig:screenshot003}
\end{figure}

 		\newtheorem{qm4}[subsection]{Lemma}
 \begin{qm4}
 	
 $	\Psi (t,x)=e^{-nit}\prod_{j\le n}  I_{x_j}(it)$ is a solution to the Schrödinger equation	$\hat{H}\Psi(t,x)=-\frac{1}{2}\sum_i(\Psi(t, x+e_i)+\Psi(t,x-e_i)-2\Psi(t,x))$ with the initial wavefunction $\Psi(0)$ being one at point zero and zero elsewhere.
 
 \end{qm4}

\paragraph{Proof}
For brevity I will prove the two-dimensional case $\Psi(t, x,y) = e^{-2it} I_x(it) I_y(it)$ (39), but the general $n$-dimensional case (38) can be proved in the same way. In this case, 	$i \frac{\partial}{\partial t} \Psi(t, x,y)=$
\begin{align}
 \frac{1}{2} (4 \Psi(t,x,y) - \Psi(t, x+1,y) - \Psi(t, x-1,y,) - \Psi(t, x,y+1) - \Psi(t, x,y-1)).
\end{align}
First, we check that the proposed solution (39) satisfies the initial condition. We can check that it follows from the definition of the modified Bessel function $I$ that $I_x(0)$ is one at $x=0$ and zero elsewhere (Olver and Maximon [2010], §10.25). So, at $t=0$,  $e^{-2it} I_x(it) I_y(it)$ is indeed equal to one at $x=y=0$ and zero elsewhere. 

We know that the modified Bessel function satisfies the following equation (Olver and Maximon [2010], §10.29):
\begin{align}
	2 \frac{\partial I_x(it)}{\partial t} = i I_{x-1}(it) + i I_{x+1}(it).
\end{align}
By differentiating  $\Psi(t, x,y) = e^{-2it} I_x(it) I_y(it)$ and making use of (41), we get: 
\begin{align}
	i \frac{\partial \Psi(t, x,y)}{\partial t} & = 2 e^{-2it} I_x(it) I_y(it) + i e^{-2it} I_x(it) \frac{\partial I_y(it)}{\partial t} + i e^{-2it} \frac{\partial I_x(it)}{\partial t} I_y(it) \\
	& = 2 \Psi(t, x,y) - \frac{1}{2} \Psi(t, x,y-1) - \frac{1}{2} \Psi(t, x,y+1) - \frac{1}{2} \Psi(t, x-1,y) - \frac{1}{2} \Psi(t, x+1,y).
\end{align}
 (43) matches up with (40). Lemma B.4 follows. $\square$

	\newpage 
	
	\section*{Acknowledgements}
	
I am greatly indebted to Tobias Fritz for his crucial guidance on the mathematical proofs contained in this paper, the codes for the figures, as well as many helpful discussions. I thank the referees at \textit{British Journal of the Philosophy of Science} for their help in improving the details of the paper. Thanks to Chris Meacham for his helpful feedback on an early draft. Many thanks to my commentators David John Baker and Jesse Fitts as well as all the attendants respectively at the 2022 APA pacific division meeting and at the Umass 16th Biennial Homecoming Conference (in honour of the retirement of Phillip Bricker)  for the intriguing discussions. This paper would also have been impossible without the past discussions with Phillip Bricker and Jeffrey Russell.
	
	\begin{flushright}
		\emph{
			Lu Chen\\
			Koc University\\
			Philosophy Department\\
		 İstanbul Türkiye	\\
		luchen@ku.edu.tr
		}
	\end{flushright}

\end{document}